\documentclass[a4paper,twocolumn,11pt,amsmath,amssymb,superscriptaddress,accepted=2022-11-02,]{quantumarticle}

\pdfoutput=1
\usepackage[utf8]{inputenc}
\usepackage[english]{babel}
\usepackage[T1]{fontenc}
\usepackage{amsmath}
\usepackage{hyperref}
\usepackage{leftidx}
\usepackage{enumitem}
\usepackage[numbers,sort&compress]{natbib}

\def\braket#1#2{\left\langle{{#1}}\mathrel{\left|{\vphantom{{#1}{#2}}}\right.\kern-\nulldelimiterspace}{{#2}}\right\rangle}

\begin{document}

\title{Entanglement-enhanced test proposal for local Lorentz-symmetry violation via spinor atoms}

\author{Min Zhuang}
\affiliation{College of Physics and Optoelectronic Engineering, Shenzhen University, Shenzhen 518060, China}

\author{Jiahao Huang}
\email{Email: hjiahao@mail2.sysu.edu.cn, eqjiahao@gmail.com}

\affiliation{Guangdong Provincial Key Laboratory of Quantum Metrology and Sensing $\&$ School of Physics and Astronomy, Sun Yat-Sen University (Zhuhai Campus), Zhuhai 519082, China}
\affiliation{State Key Laboratory of Optoelectronic Materials and Technologies, Sun Yat-Sen University (Guangzhou Campus), Guangzhou 510275, China}
\author{Chaohong Lee}
\email{Email: chleecn@szu.edu.cn, chleecn@gmail.com}
\affiliation{College of Physics and Optoelectronic Engineering, Shenzhen University, Shenzhen 518060, China}
\affiliation{Guangdong Provincial Key Laboratory of Quantum Metrology and Sensing $\&$ School of Physics and Astronomy, Sun Yat-Sen University (Zhuhai Campus), Zhuhai 519082, China}
\affiliation{State Key Laboratory of Optoelectronic Materials and Technologies, Sun Yat-Sen University (Guangzhou Campus), Guangzhou 510275, China}
\maketitle

\begin{abstract}
  Invariance under Lorentz transformations is fundamental to both the standard model and general relativity.
  Testing Lorentz-symmetry violation (LSV) via atomic systems attracts extensive interests in both theory and experiment.
  In several test proposals, the LSV violation effects are described as a local interaction and the corresponding test precision can asymptotically reach the Heisenberg limit via increasing quantum Fisher information (QFI), but the limited resolution of collective observables prevents the detection of large QFI.
  Here, we propose a multimode many-body quantum interferometry for testing the LSV parameter $\kappa$ via an ensemble of spinor atoms.
  By employing an $N$-atom multimode GHZ state, the test precision can attain the Heisenberg limit $\Delta \kappa \propto 1/(F^2N)$ with the spin length $F$ and the atom number $N$.
  We find a realistic observable (i.e. practical measurement process) to achieve the ultimate precision and analyze the LSV test via an experimentally accessible three-mode interferometry with Bose condensed spin-$1$ atoms for example.
  By selecting suitable input states and unitary recombination operation, the LSV parameter $\kappa$ can be extracted via realizable population measurement.
  Especially, the measurement precision of the LSV parameter $\kappa$ can beat the standard quantum limit and even approach the Heisenberg limit via spin mixing dynamics or driving through quantum phase transitions.
  Moreover,the scheme is robust against nonadiabatic effect and detection noise.
  Our test scheme may open up a feasible way for a drastic improvement of the LSV tests with atomic systems and provide an alternative application of multi-particle entangled states.
\end{abstract}

\section{Introduction}
Local Lorentz-symmetry violation is one of the most fundamental principles of modern physics and a key feature of the standard model~\cite{Ref1,Ref2,Ref3,Ref4,Ref5}.
It means that the outcome of any local non-gravitational experiment is independent of the velocity and the orientation of the apparatus.
However, some theories aim at unifying gravity with other fundamental interactions, including string theory and field theory, imply the possibility or even necessity for Lorentz-symmetry violation (LSV)~\cite{Ref6,Ref7,Ref8,Ref9,Ref10,Ref11,Ref12,Ref13,PRL94081601,Ref15,Ref16,Ref17,Ref18}.
The high-precision tests of LSV with matter, gravity, or light have the potential to offer insight about new physics and set limits on various theories.
The effects of LSV can be classified in the framework of the standard model extension (SME)~\cite{Ref7,Ref10,Ref9,Ref22}.
The SME is an effective field theory that augments the standard model Lagrangian with all combination of the standard model fields that are not term-by-term invariant under Lorentz transformations, but maintain gauge invariance, energy-momentum conservation, and Lorentz invariance of the total action.

Based on the SME, the LSV tests with atomic systems have attracted vast attention~\cite{RMP90025008}.
The LSV tests with electronic states have conducted via neutral Dy atoms~\cite{Ref23}, Ca$^{+}$ ions~\cite{Ref24} and a pair of two entangled trapped Yb$^{+}$ ions~\cite{Ref25}.
Recently, it was proposed to test LSV via decoherence free subspace (DFS)~\cite{Ref25} or dynamical decoupling~\cite{Ref26} to cancel out the influence of magnetic field fluctuations.
For an atom with fixed total angular momentum $F$, the LSV in bound electronic states result in a small shift of the energy levels~\cite{PRD601160101999,Ref23,Ref24,Ref25,Ref26,Ref29,Ref31}, one energy level shift term is proportional to the square of the $z$ component of the total angular momentum~\cite{PRD601160101999,Ref26,Ref31}, i.e., $\Delta E_{LSV}\propto \kappa \hat{s}_{z}^{2} $ with the LSV parameter $\kappa$, which is analogous to quadratic Zeeman shift.

It is well known that quantum multi-particle entanglement can implement high-precision measurement beyond the standard quantum limit (SQL)~\cite{Ref32,Ref33,Ref34,Ref35,PRL971504022006,PRL1020704012009,Ref38}.
Recently, utilizing multi-particle entangled state to enhance test precision of LSV has been proposed~\cite{Ref31}.
The study indicates that when ensembles of entangled particles are employed~\cite{Ref31}, the QFI of LSV parameter $\kappa$, an equivalent quadratic Zeeman shift can asymptotically approach the Heisenberg limit~\cite{Ref32,Ref33,Ref34,Ref35,PRL971504022006,PRL1020704012009}.
The QFI provides an ultimate bound for the test precision of LSV via multipartite entanglement, but it is very important to find a realistic observable or practical measurement process to achieve the ultimate precision.

Bose-condensed spinor atoms involve multiple spin degrees of freedom~\cite{Ref39}, can serve as an excellent candidate for implementing multimode quantum interferometry to achieve high-precision measurement~\cite{Ref39,Ref40}.
In the presence of magnetic field, the hyperfine states $|F,m_F\rangle$ are split separately and then can act as multiple paths or multiple modes.
Naturally, the following questions arise: i) can one use multimode many-body quantum states to test the LSV parameter $\kappa$ ?
ii) what is the ultimate measurement precision bound? and
iii) is it possible to implement in a state-of-the-art experiment and find a realistic observable?

In this paper, by considering an ensemble of Bose condensed spinor atoms, we propose a multimode many-body quantum interferometry for estimating the LSV parameter $\kappa$.
We find that, if the atoms are prepared in a multimode GHZ state, the measurement precision of $\kappa$ can attain the Heisenberg limit, and it depends on the single-atom spin length $F$ according to $\Delta \kappa \propto 1/(F^2N)$.
Further, we take an experimentally accessible three-mode interferometry with spin-$1$ Bose-condensed atoms for example and show how to achieve the sub-SQL measurement precision via population measurement without request of single-particle resolution~\cite{PRL971504022006,Burd,Linnemann2016,Hosten2016,Mirkhalaf2018,FFrowis2016,Szigeti2017,JHuang2018,Anders2018,JHuang97032116}.
Our test scheme may open up a feasible way to achieve high-precision estimation of the LSV parameter $\kappa$.
Given the broad interest in testing LSV and quantum-enhanced metrology, our study not only may pave the way to test some theories, such as string theory~\cite{Ref8}, field theory~\cite{Ref6,Ref12,Ref9,Ref7}, Einstein-aether theories~\cite{arXiv08011547} and quantum gravity~\cite{PRL104181302}, but also provides an alternative application of multi-particle entangled states~\cite{Ungar2020}.
Meanwhile, the energy shift resulted from LSV is similar to the quadratic Zeeman shifts in atomic clocks~\cite{Metrologia42411,Metrologia4982,JPB47015001} and atomic magnetometry~\cite{PhysScrT10527,PRA75051407,PRA79023406}, thus our test proposal also can be used in practical quantum sensors.

The paper is organized as follows.
In Sec.~\ref{Sec2}, we introduce the scheme of multimode interferometry via spin-$F$ Bose-condensed atoms for testing the LSV parameter $\kappa$ .
In Sec.~\ref{Sec3}, we discuss the ultimate precision bound for the LSV parameter $\kappa$ via individual atoms as well as entangled atoms.
In particular, using a specific $N$-atom multimode GHZ state as an input state, the measurement precision of the LSV parameter $\kappa$ can reach the Heisenberg limit.
In Sec.~\ref{Sec4}, as an example, we propose an experimentally accessible three-path interferometric scheme via spin-$1$ Bose-condensed atoms and show how to push the test precision beyond SQL via population measurement.
Finally, we give a brief summary in Sec.~\ref{Sec5}.

\section{LSV test via quantum interferometry of spinor atoms}\label{Sec2}

Invariance under Lorentz transformations is fundamental in modern physics.
In relativistic physics, Lorentz symmetry implies the laws of physics stay the same for all observers moving at constant velocities with respect to each other in inertial frames.
It is also described as the outcome of any local experiment is independent of the velocity and the orientation of the (freely falling) apparatus.
However, the Lorentz symmetry might be violated at experimentally accessible energy scales due to spontaneous symmetry breaking~\cite{Ref15}.
In addition, some theories that unify gravitation and the standard model assert that Lorentz symmetry is valid only at large length scales~\cite{PRL94081601}.
For atomic systems, the effect of various sources of LSV on atoms has been discussed in Ref.~\cite{Ref22}.
Within the framework of SME, the QED Lagrangian (for electron) for an electron sector with LSV can be written as~\cite{Ref9,Ref10,Ref24}
\begin{equation}\label{Lagranian}
L=\frac{1}{2}i \bar{\psi}(\gamma_{\nu}+c_{\mu\nu}\gamma^{\mu})\stackrel{\leftrightarrow}{D}
\!^{\nu} {\psi}-\bar{\psi}m_{e}{\psi},
\end{equation}
where $m_e$ is the electron mass, $\psi$ is a Dirac spinor, $\gamma^{\mu}$ are the Dirac matrices and $\bar{\psi}\!\stackrel{\leftrightarrow}{D}^{\nu}\!{\psi}\equiv \bar{\psi}{D}^{\nu}{\psi}-{\psi}{D}^{\nu}\bar{\psi}$, with ${D}^{\nu}$ being the covariant derivative.
The tensor $c_{\mu\nu}$ in the above Eq.~(\ref{Lagranian}) quantifies the strength of LSV by the frame dependent interaction term~\cite{Ref6,Ref7,Ref9}.
According to Eq.~(\ref{Lagranian}), the LSV in bound electronic states results in a small energy level shift described by the Hamiltonian~\cite{PRD601160101999,Ref23, Ref24}:
\begin{equation}\label{Lagranian1}
\delta H_{LSV}= -(C_{0}^{(0)}-\frac{2U}{3c^2}c_{00})\frac{\emph{\textbf{p}}^{2}}{2}-\frac{1}{6}C_{0}^{(2)}T_{0}^{(2)}.
\end{equation}
Here, $U$ is the Newtonian gravitational potential, $\emph{\textbf{p}}$ is the momentum of a bound electron, and $c$ is the speed of light.
The more specific parameters $C_{0}^{(0)}$, $c_{00}$ and $C_{0}^{(2)}$ contain elements of $c_{\mu\nu}$ tensor, thus they can quantify the strength of LSV~\cite{Ref24}.
We just consider the second term of Eq.~(\ref{Lagranian1}) which describe the dependence of electron's energy on the distribution of the total momentum among the three spatial components.
The non-relativistic form of the $T_{0}^{(2)}$ operator is $T_{0}^{(2)}=(\emph{\textbf{p}}^{2}-3p_{z}^{2})/m_{e}$ with the electron mass $m_{e}$ and the component of the electron's momentum $p_{z}$ along the quantization axis fixed in the laboratory frame.
%
Thus, for a bound electron system with momentum $\emph{\textbf{p}}$, the non-relativistic form of the second term of $\delta H_{LSV}$  reduces to
 ~\cite{PRD601160101999,Ref23,Ref24,Ref25,Ref26,Ref29}
\begin{equation}\label{Eq1}
{\hat{h}}=-C_{0}^{(2)} \frac{\emph{\textbf{p}}^{2}-3p_{z}^{2}}{6m_{e}}.
\end{equation}
%
%
The Hamiltonian Eq.~(\ref{Eq1}) is responsible for breaking the symmetry of orientation or velocity in a bound electron system.
It illustrates that the energy shift depends on the distribution of total momentum $\emph{\textbf{p}}$ among the three spatial components.
The $c_{\mu \nu}$ tensor is frame-dependent, thus $C_{0}^{(2)}$ varies in time as the Earth rotates, resulting in a time variation of the electron's energy correlated with the Earth's motion.

Now, we consider an ensemble of spinor Bose condensed atoms with orbital angular momentum $L=0$ and spin angular momentum $F$, thus the total angular momentum of the spinor atom is $F$.
In the presence of magnetic field, there are $2F+1$ Zeeman sublevels denoted by $|{F,m_F}\rangle$, with $m_F=-F,-F+1,...,F$ being the magnetic quantum number.
According to the Wigner-Eckart theorem, the matrix element of operator $T_{0}^{(2)}$ can be expressed through the reduced matrix element $\langle F|T^{(2)}|F\rangle$ and it is~\cite{Ref31}
\begin{eqnarray}\label{Eq2}
 &&\langle F,m_F|\frac{\emph{\textbf{p}}^{2}\!-\!3p_{z}^{2}}{6m_{e}}|F,m_F\rangle\nonumber\\
 &&= \frac{[-F(F+1)\!+\!3m_{F}^{2}]\langle F|T^{(2)}|F\rangle}{\sqrt{(2F+3)(F+1)(2F\!+\!1)(2F-1)}}.\nonumber\\
\end{eqnarray}
%
%
%
According to Eq.~(\ref{Eq2}), the LSV signal contains a term proportional to $m_F^2$, thus the LSV dynamics can be described by an equivalent Hamiltonian $\hat{h}^{{L}} = \kappa \hat{s}_{z}^2$ with the magnetic angular momentum operator $\hat{s}_z$~\cite{Ref26,Ref31}.
The equivalent Hamiltonian is analogous to a quadratic Zeeman shift.
Thus, the LSV dynamics can be described by the equivalent Hamiltonian $\hat{h}^{{L}}$, and
the goal of placing a bound on the parameter $C_{0}^{(2)}$ is equivalent to placing a bound on the parameter $\kappa$.

In the following, we describe the LSV dynamics via the equivalent Hamiltonian $\hat{h}^{{L}}$.
For $N$ spin-$F$ atoms, the LSV Hamiltonian can be expressed as
\begin{eqnarray}\label{Eq3}
 \hat{H}^{{L}}_{F} = \sum\limits_{n=1}^N \hat{h}^{{L} [n]} =\kappa \sum\limits_{n=1}^N \hat{s}^{2 [n]}_z=\kappa \hat{G},
\end{eqnarray}
where $n$ denotes the $n$-th atom with $n=1,2,\ldots,N$.
The procedures of multimode quantum interferometer are as follows.
In the initialization stage, an $N$-particle input state $|{\Psi_{{in}}}\rangle$ involving $2F+1$ Zeeman sublevels is prepared.
Then, the multimode input probe state undergoes the LSV dynamics and acquires a phase shift.
The information of the LSV parameter $\kappa$ is imprinted in the evolved output state $|\Psi_{{{out}}}\rangle={e}^{-i \hat{H}^{{L}}_{F} T}|\Psi_{{in}}\rangle$ with the evolution time $T$.
In the readout stage, the unknown parameter $\kappa$ can be extracted by selecting suitable measurement on the output state.
\section{Measurement precision bounds}\label{Sec3}
In this section, we discuss the ultimate measurement precision bound for the LSV parameter $\kappa$ via our multimode interferometer with different input states.
Without loss of generality, we assume the evolution time $T=1$ in our calculation.
In Sec.~\ref{Sec3a} and Sec.~\ref{Sec3b}, we show the ultimate measurement precision bound of $\kappa$ via individual and entangled Bose condensed spin-$F$ atoms, respectively.

According to the parameter quantum estimation theory~\cite{PRL7234391994,GTOth2014,RDemkowicz2015,PRL1021004012009,PRL1051205012010,Annu2365},
the ultimate measurement precision of $\kappa$ is determined by the quantum Cram\'{e}r-Rao bound(QCRB) and it is
\begin{eqnarray}\label{Eq5}
  \Delta \kappa \geq \Delta \kappa_{QCRB}\equiv \frac{1}{\sqrt{\eta F_{Q}}}.
\end{eqnarray}
Here, $\eta$ corresponds to the number of trials, $F_{Q}$ is the quantum Fisher information(QFI) and it can be written as
\begin{eqnarray}\label{Eq6}
  F_{Q}=4(\langle \Psi '|\Psi '\rangle-|\langle \Psi '|\Psi_{{out}}\rangle|^2)=4 \Delta^2 \hat{G},
\end{eqnarray}
where $|{\Psi'}\rangle=d|\Psi_{{out}}\rangle/d \kappa$ denotes the derivative of the output state respective to the parameter $\kappa$, and $\Delta^2 \hat{G}=\langle \Psi_{{in}}|\hat{G}^{2}|\Psi_{{in}}\rangle-(\langle \Psi_{{in}}|\hat{G}|\Psi_{{in}}\rangle)^2$ is the variance of $\hat{G}$.
\subsection{Measurement precision bounds offered by individual atoms \label{Sec3a}}
%
%
We first consider individual atoms without any entanglement.
Thus, the system state can be described by a product state of $N$ atoms, which is written as~\cite{Ref40}
\begin{equation}\label{Eq7}
 |{\Psi}\rangle^{{Pro}}=\left(\sum\limits_{m_F=-F}^{F}\!\!\alpha_{m_F}|{F,m_F}\rangle\right)^{\otimes N},
\end{equation}
where $\alpha_{m_F}$ is the complex amplitude and satisfies the normalization condition:
$\sum\limits_{m_{F}=-F}^{F}\!\!|\alpha_{m_{F}}|^2=1$.
Taking the $N$-atom product state~(\ref{Eq7}) as the input state, the ultimate bound of $\kappa$ is ${\Delta^{2} \kappa}_{QCRB}=\frac{1}{4N(M_2-M_1^2)}$, with $M_1=\sum_{{m_F}=-F}^{F}|\alpha_{{m_F}}|^2 m_F^2$ and $M_2=\sum_{{m_F}=-F}^{F}|\alpha_{{m_F}}|^2 m_F^4$.
It is obvious that the QFI only scales linearly with $N$.

If the input state is in uniform distribution among the Zeeman sublevels~\cite{Ref40},
the corresponding ultimate measurement precision bound for parameter $\kappa$ is ${\Delta \kappa}_{{QCRB}}=\sqrt{\frac{45}{4NF(1+F)(4F^2+4F-3)}}$.
Further, the optimal state for individual atoms can be obtained via minimizing the ultimate measurement precision bound~\cite{Ref40} and the corresponding ultimate measurement precision bound is ${\Delta \kappa}_{{QCRB}}=\frac{1}{\sqrt{N}F^{2}}$.
%
\subsection{Measurement precision bounds offered by entangled atoms}\label{Sec3b}
As we have known, entanglement is a useful quantum resource to improve the measurement precision~\cite{Annu2365,PRL7234391994,VGiovannetti2004,VGiovannetti2011,Bohnet2016}.
In the following, we show how the entangled input states of $N$ atoms can enhance the measurement precision of the LSV parameter $\kappa$.
In particular, we consider a multimode GHZ state as an input state and the form is~\cite{Ref40}
\begin{equation}\label{Eq17}
  |{\Psi}\rangle^{{GHZ}}=\sum\limits_{{m_F}=-F}^{F}\alpha_{{m_F}}|{F,{m_F}}\rangle^{\otimes N}.
\end{equation}
Substituting the input state $|{\Psi}\rangle^{{GHZ}}$ into Eq.~(\ref{Eq5}), the corresponding ultimate precision of $\kappa$ is ${\Delta^{2}{\kappa}}_{{QCRB}}=\frac{1}{4N^2(M_2-M_1^2)}$.
Compared with individual atoms, the scaling of the ultimate measurement precision bound versus total atom number $N$ changes from $1/\sqrt{N}$ to $1/N$.
It is evident that the GHZ's form of $N$-atom state can improve the measurement precision of the LSV parameter $\kappa$ to the Heisenberg limit.
For the multimode GHZ state, the optimal one reads as $|{\Psi}\rangle_{{opt}}^{{GHZ}}=\frac{1}{2}|{F,F}\rangle^{\otimes N}+\frac{1}{\sqrt{2}}|{F,0}\rangle^{\otimes N}\!+\frac{1}{2}|{F,-F}\rangle^{\otimes N}$
and the corresponding ultimate measurement precision bound for parameter $\kappa$ is ${\Delta \kappa}_{{QCRB}}=\frac{1}{N F^{2}}$.

The ultimate measurement precision of the LSV parameter $\kappa$ decreases monotonously as $F$ increases for individual or entangled atoms, thus one can improve the measurement precision of $\kappa$ by using the atoms with larger spin $F$~\cite{Ref40}.

\section{LSV test via three-mode interferometry with Bose-condensed spin-1 atoms\label{Sec4}}
In experiments, how to prepare the desired input states for sensing is an important problem.
Despite multimode GHZ states can make measurement precision attain the Heisenberg limit in theory, they are not easy to be generated in experiments.
Meanwhile, to perform the interferometry with GHZ states in practice, parity measurement is required~\cite{Ref32,Ref34}.
However, parity measurement requires single-particle resolution technique and limits the experimental feasibility of quantum metrology with GHZ states.
How to prepare an available entangled state and beat the SQL without single-particle-resolved detection become essential.
\begin{figure}[!htp]
\includegraphics[width=\columnwidth]{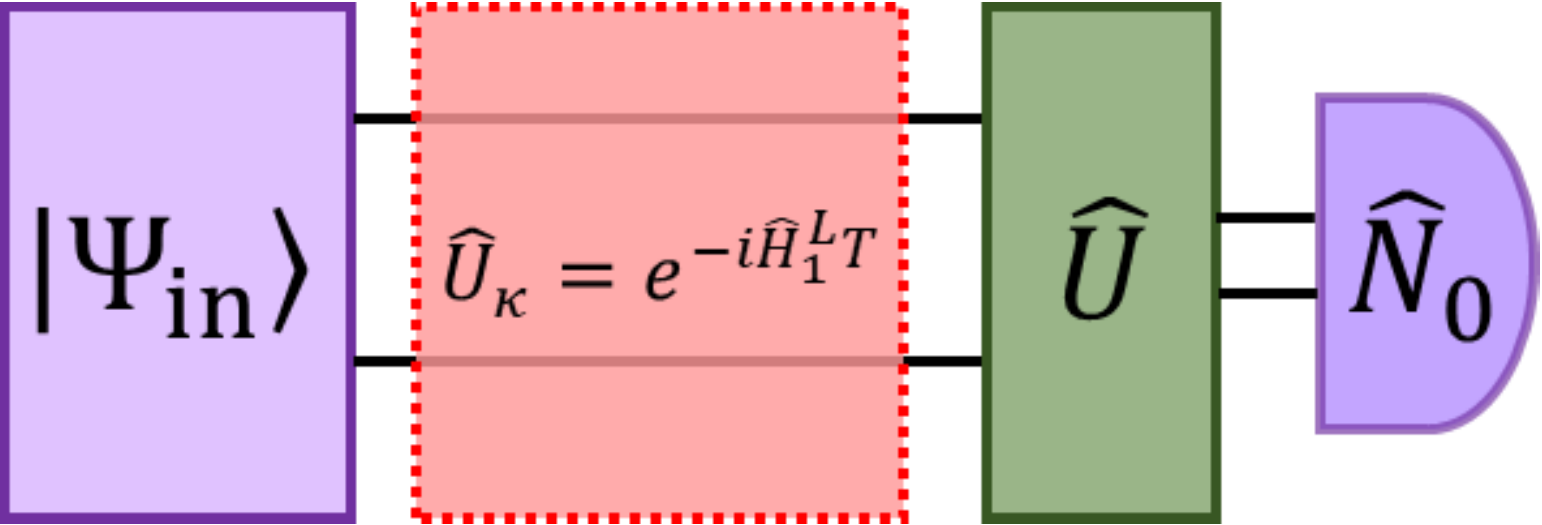}
\caption{\label{Fig3}(color online).
Schematic of three-mode interferometry. Firstly, a suitable input state is prepared. Secondly, interrogation for accumulating phase is implemented. Thirdly, an unitary operation $\hat{U}$ is applied for recombination. Lastly, the population measurement $\hat{N}_{0}$ is applied to extract the parameter $\kappa$}.
\end{figure}
In this section, we discuss the test of LSV via three-mode interferometry with spin-$1$ Bose-condensed atoms~\cite{ZZhang2013,YZou2018,Luo2017,SGuo2021}.
For spin-$1$ Bose-condensed atoms, the LSV dynamics can be written as
\begin{equation}\label{Eq21}
 \hat{H}_{F=1}^{L}=\kappa (\hat{N}_{-1}+\hat{N}_{1}).
\end{equation}
Here $\hat{N}_{m_{F}}=\hat{a}^{\dagger}_{m_{F}}\hat{a}_{m_{F}}$ denotes the particle number operator of atoms in state $|{1,{m_{F}}}\rangle$, with $\hat{a}^{\dagger}_{m_{F}}$ and $\hat{a}_{m_{F}}$ being the creation and annihilation operators for state $|{1,{m_{F}}}\rangle$, respectively.
For convenience, we abbreviate the Hamiltonian of LSV dynamics $\hat{H}_{F=1}^{L}$ to $\hat{H}_{1}^{L}$, i.e., $\hat{H}_{1}^{L}=\kappa (\hat{N}_{-1}+\hat{N}_{1})$.
Now, we illustrate the procedures of three-mode interferometry in detail (see Fig.~\ref{Fig3}).
Firstly, the system is initialized in a desired input state $|{\Psi_{{in}}}\rangle$.
Then, the input state $|{\Psi_{{in}}}\rangle$ undergoes the LSV dynamics and a phase dependent on $\kappa$ is accumulated .
The output state $|{\Psi_{{out}}}\rangle=e^{-i\hat{H}_{1}^{L}T}|{\Psi_{{in}}}\rangle$ contains the information of the parameter $\kappa$.
Finally, a suitable unitary operation $\hat{U}$ is performed on the output state $|{\Psi_{{out}}}\rangle$ for recombination and the final state can be written as
\begin{equation}\label{Eq22}
|{\Psi_{final}}\rangle=\hat{U}e^{-i\hat{H}_{1}^{L}T}|{\Psi_{{in}}}\rangle.
\end{equation}
After applying the population measurement $\hat{N}_0$ on the final state, one can obtain the expectation and standard deviation of $\hat{N}_0$,
\begin{eqnarray}\label{Eq23}
\langle \hat{N}_{0} \rangle
=\langle{\Psi_{{{final}}}}| \hat{N}_{0} |{\Psi_{{final}}}\rangle  \nonumber \\
\end{eqnarray}
\begin{eqnarray}\label{Eq23}
\Delta{\hat{N}_{0}}\!=\!\sqrt{\langle{\Psi_{{final}}}| \hat{N}_{0}^2 |{\Psi_{{final}}}\rangle\!-\!(\langle{\Psi_{{{final}}}}| \hat{N}_{0} |{\Psi_{{final}}}\rangle)^{2}}.\nonumber \\
\end{eqnarray}
According to the error propagation formula, the measurement precision of the estimated parameter $\kappa$ read as
\begin{eqnarray}\label{Eq24}
\Delta \kappa=\frac{\Delta{\hat{N}_{0}}}{|\partial{\langle\hat{N}_{0}\rangle}/ \partial{\kappa}|}.
\end{eqnarray}

For spin-$1$ Bose condensed atoms, spin exchange collision can be used to generate different kinds of input states.
In the following, we study how to realize the high-precision measurement for the LSV parameter $\kappa$ via spin-mixing dynamics (SMD) and driving through quantum phase transitions (QPTs), and study its measurement precision via population measurement.
%
\subsection{LSV test via spin-mixing dynamics\label{Sec4a}}
\begin{figure}[!htp]
  \includegraphics[width=1\columnwidth]{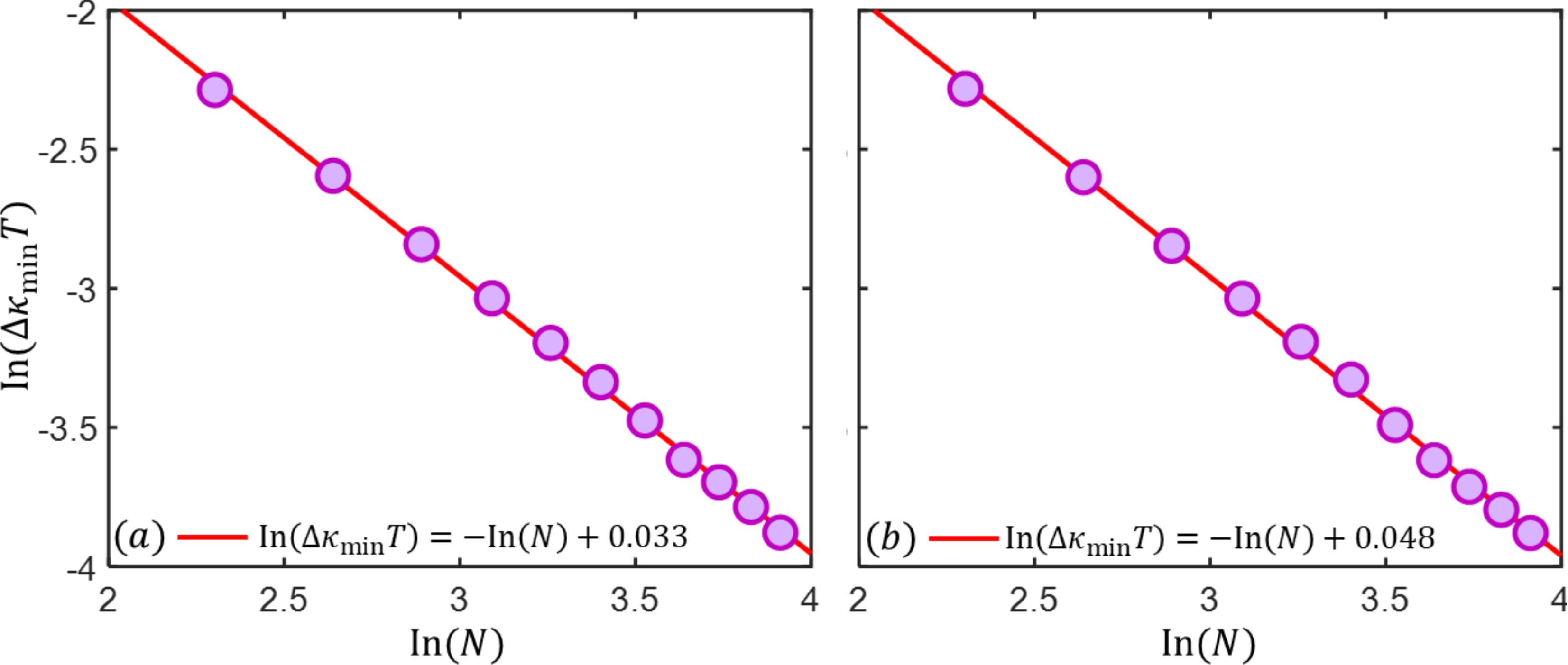}
  \caption{\label{Fig4}(color online).
  The log-log scaling of $\Delta \kappa_{min}$ versus the total atom number $N$ under SMD.
  (a) The circles show results for the input state $|{\Psi_{{in}}}\rangle=e^{-i\hat{H}_{SMD}t}|{0,N,0}\rangle$;
  (b) The circles show results for the input state $|{\Psi_{{in}}}\rangle=(|{0,N,0}\rangle+|{N/2,0,N/2}\rangle)/\sqrt{2}$.
  Here, $\kappa T\in [-0.1\pi,0.1 \pi]$, $t\in (0,2 \pi]$, $\chi =1$, $T=1$}
\end{figure}
The input state $|{\Psi_{{in}}}\rangle$ can be generated by time evolution under spin-mixing dynamics~\cite{TLHo1998,TOhmi1998,DMStamperKurn2013,MGabbrielli2015}.
The governed Hamiltonian is written as
\begin{equation}\label{Eq27}
  \hat{H}_{SMD}=\chi \left(\hat{a}^{\dagger}_{0}\hat{a}^{\dagger}_{0}\hat{a}_{1}\hat{a}_{-1}
  + \hat{a}_{0}\hat{a}_{0}\hat{a}^{\dagger}_{1}\hat{a}^{\dagger}_{-1}\right),
\end{equation}
where $\chi $ is the strength of spin exchange collision.
Initially with all $N$ atoms in state $|{1,{m_{F}=0}}\rangle$, i.e., $|{\Psi_0}\rangle=|{0,N,0}\rangle$.
Then, the input state can be generated via SMD,
\begin{equation}\label{Eq28}
  |{\Psi_{{in}}}\rangle=e^{-i\hat{H}_{SMD}t}|{\Psi_0}\rangle=e^{-i\hat{H}_{SMD}t}|{0,N,0}\rangle,
\end{equation}
where $t$ is the evolution time.
For a fixed $\chi$, different evolution time can result in different input state.
Especially, the input state satisfies $N_{1}=N_{-1}$, thus it is immune to stray magnetic fields and forms a DFS~\cite{Ref25,Ref31}.
For a fixed total atom number $N$, one can numerically obtain all the input state during the time evolution.
Here, we choose $\hat{U}=e^{i\hat{H}_{SMD}t}$ to realize the recombination, which can be regarded as a nonlinear detection~\cite{Mirkhalaf2018,EDavis2016,TMacri2016,JHuang97032116,FFrowis2016,Szigeti2017,JHuang2018,Anders2018,SPNolan2017,LucaPezze2013}.
The final state can be written as
\begin{eqnarray}\label{Eq29}
  |{\Psi_{{final}}}\rangle&=&e^{i\hat{H}_{SMD}t}e^{-i\hat{H}^{{L}}_{1}T}e^{-i\hat{H}_{SMD}t}|{\Psi_0}\rangle \nonumber \\
                          &=&e^{i\hat{H}_{SMD}t}e^{-i\kappa (\hat{N}_{-1}+\hat{N}_{1})T}e^{-i\hat{H}_{SMD}t}|{\Psi_0}\rangle. \nonumber \\
\end{eqnarray}
Finally, one can extract the estimated parameter $\kappa$ via measuring the population $\hat{N}_{0}$ on the final state.
%
%

According to Eq.~(\ref{Eq24}), we obtain the corresponding $\Delta \kappa$ via population measurement.
Minimizing $\Delta \kappa$, we can obtain the optimal input state generated via SMD.
Although we cannot give the analytical form of the optimal input state for a given $N$, we numerically confirm that SMD could be an effective way for preparing the suitable states for estimation.
To confirm the dependence of measurement precision $\Delta \kappa$ on the total particle number $N$, we numerically calculate
the optimal measurement precision versus particle number in the range of $t\in [0,2\pi] $ and $\kappa T \in [-0.1\pi,0.1\pi] $, as shown in Fig.~\ref{Fig4}(a).
According to the fitting results, the optimal measurement precision is given by $\textrm{{In}}(\Delta \kappa_{{min}}T)=-\textrm{In}(N)+0.033$.

Further, if the atoms are prepared in state $|{\Psi_{{in}}}\rangle=\frac{1}{\sqrt{2}}\left(|{0,N,0}\rangle+|{N/2,0, N/2}\rangle\right)$, the corresponding final state is
$|{\Psi_{{final}}}\rangle=e^{i\hat{H}_{SMD}t}e^{-i\hat{H}^{{L}}_{1}T}|{\Psi_{{in}}}\rangle$.
%
Similarly, we numerically calculate the optimal measurement precision versus particle number in the range of $t\in [0,2\pi]$ and $\kappa T\in [-0.1\pi,0.1\pi]$ , as shown in Fig.~\ref{Fig4}(b).
According to the fitting result, the optimal measurement precision is given by $\textrm{{In}}(\Delta \kappa_{{min}}T)=-\textrm{{In}}(N)+0.048$.
%
Our results indicate that SMD could be an effective way for states preparation and beam recombination for measuring the LSV parameter $\kappa$,  the measurement precision $\Delta \kappa$ can beyond SQL and exhibit the Heisenberg limit via population measurement.
\subsection{LSV test via driving through quantum phase transitions\label{Sec4b}}
\begin{figure}[!htp]
  \includegraphics[width=1\columnwidth]{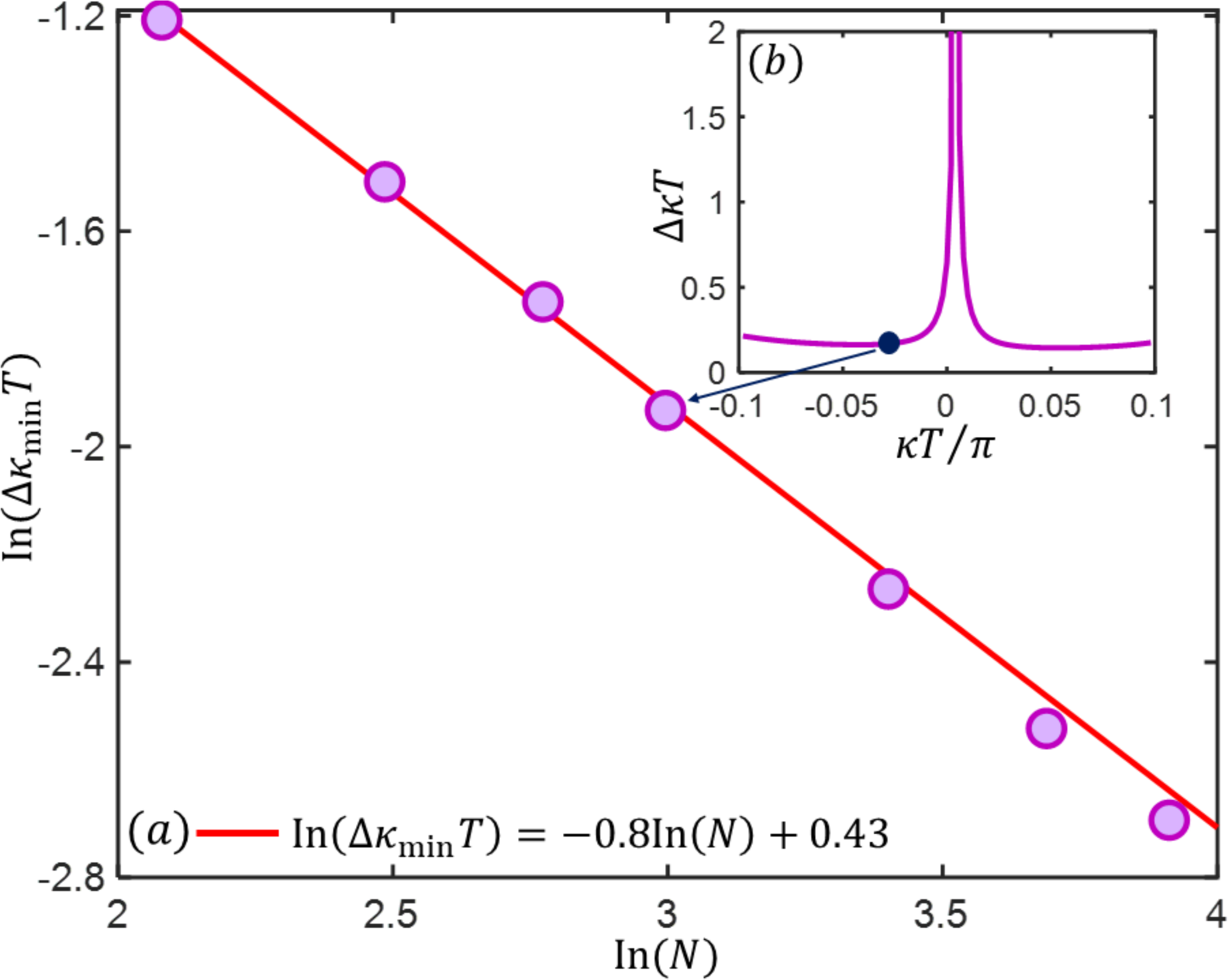}
  \caption{\label{Fig5}(color online).
  The measurement precision via population measurement.
  (a)The log-log scaling of $\Delta \kappa_{min}T$ versus the total atom number $N$ under driving through QPTs.
  (b) Variation of measurement precision versus $\kappa$ under $N=20$.
  Here, $\kappa T\in [-0.1\pi,0.1 \pi]$, $q_{0}=q_{f}=3$, $\beta=0.01$, $T=1$}
\end{figure}
For Bose-condensed spin-$1$ atoms, an alternative way for preparing entangled input state is to drive the system through QPTs~\cite{ZZhang2013,YZou2018,Luo2017,SGuo2021}.
Despite SMD can generate the desired input state, it sensitively depends on the control of evolution time and these states are always not steady.
In contrast, driving system through QPTs can deterministically generate the entangled states~\cite{Ref34,PRL971504022006,PRL1020704012009,ZZhang2013,YZou2018,Luo2017,SGuo2021,PRA930436152016,Ref61}.
Recently, entanglement generation by driving through QPTs has been realized in $^{87}Rb$ spinor condensate~\cite{YZou2018,Luo2017,SGuo2021}.
The evolution of the initial state $|{\Psi_0}\rangle=|{0,N,0}\rangle$ is governed by the following Hamiltonian,
\begin{eqnarray}\label{Eq29}
  \hat{H}_{{{QPT}}}=&&\frac{c_2}{2N}[2(\hat{a}^{\dagger}_{0}\hat{a}^{\dagger}_{0}\hat{a}_{1}\hat{a}_{-1}
  +\hat{a}_{0}\hat{a}_{0}\hat{a}^{\dagger}_{1}\hat{a}^{\dagger}_{-1})\nonumber \\
         &&+(2\hat{N_0}-1)(N-\hat{N_0})]-q(t)\hat{N_0}.\nonumber \\
\end{eqnarray}
Here, $|c_2|$ describes the rate of spin mixing process, $q=(\varepsilon_{+1}+\varepsilon_{-1})/2-\varepsilon_{0}$, with $\varepsilon_{m_F}$ being the energy of state $|{1, m_F}\rangle$, and $q(t)$ can be tuned linearly with time in experiment.
The system possesses three distinct phases through the competition between $|c_2|$ and $q$~\cite{YZou2018,Luo2017,SGuo2021}.
For $q \gg 2|c_2|$, the ground state is a polar state with all atoms in state $|{1,0}\rangle$.
For $q \ll -2|c_2|$, the ground state becomes a twin Fock state with atoms equally populated in $|{1,-1}\rangle$ and $|{1,1}\rangle$.
When $-2|c_2| < q < 2|c_2|$, the ground state corresponds to a superposition of all three components.
If we ramp $q(t)$ from $q \gg 2|c_2|$ towards $q \ll -2|c_2|$ with very slow ramping rate, any instantaneous ground states can be adiabatically prepared.
Especially, if $q$ is adiabatically ramped from $q \gg 2|c_2|$ to zero, an initial polar state will evolve into the balanced spin-1 Dicke state $|{D}\rangle$, which can realize the measurement precision bound of the LSV parameter $\kappa$ over the SQL and asymptotically reach the Heisenberg limit~\cite{Ref31}.

Our protocol for testing LSV parameter $\kappa$ via driving through QPTs is presented below.
Firstly, we ramp $q(t) = q_0 - \beta t$ from $q \gg 2|c_2|$ towards $q=0$ adiabatically to generate the balanced spin-$1$ Dicke state  $|{D}\rangle=e^{-i \int_{0}^{\tau_1} H_{{QPT}}(t) dt}|{0,N,0}\rangle$.
Here, $\beta$ denotes sweeping rate.
Then, a rotation operation $\hat{{R}}_{\pi/2}$ is applied on the balanced spin-1 Dicke state $|{D}\rangle$ to generate the input state  $|{\Psi_{{in}}}\rangle=\hat{{R}}_{\pi/2}|{D}\rangle$ with $\hat{{R}}_{\pi/2}=e^{-i\frac{\pi}{2}\hat{L}_{x}}$ and
$\hat{L}_{x}=(\hat{a}_{1}^{\dag}\hat{a}_{0}+\hat{a}_{1}\hat{a}_{0}^{\dag}+\hat{a}_{0}^{\dag}\hat{a}_{-1}+\hat{a}_{0}\hat{a}_{-1}^{\dag})/\sqrt{2}$. 
In the interrogation step, each atom undergoes the LSV dynamics of duration $T$ and the system goes through a phase accumulation process.
In the recombination step, a second rotation operation $\hat{{R}}_{\pi/2}^{\dag}$ is applied at first and then we ramp $q(t)$ from $q=0$ towards $q \gg 2|c_2|$ adiabatically.
The corresponding unitary operation $\hat{U}$ for recombination is $\hat{U}=e^{-i \int_{\tau_1+T}^{\tau_2} H_{{QPT}}(t) dt} \hat{{R}}_{\pi/2}^{\dag}$.
The final state can be written as,
\begin{eqnarray}\label{Eq30}
 |{\Psi_{{final}}}\rangle=&&e^{-i \int_{\tau_1+T}^{\tau_2} H_{{QPT}}(t) dt} \hat{{R}}_{\pi/2}^{\dag} e^{-i\kappa (\hat{N}_{-1}+\hat{N}_{1})T}\nonumber \\ &&\hat{{R}}_{\pi/2} e^{-i \int_{0}^{\tau_1} H_{{QPT}}(t) dt}|{0,N,0}\rangle.
\end{eqnarray}
Similarly, the LSV parameter $\kappa$ can be estimated by measuring the final population in component $m = 0$.
Ideally, if $\kappa T=\frac{M\pi}{2}$ with $M$ an integer, all atoms would populate into $m = 0$ again, leading to
$\langle{\hat{N}_{0}}\rangle = N$.
For other values of $\kappa T$, the population $\langle{\hat{N}_{0}}\rangle$ would be less than $N$ and dependent on $\kappa T$.
In Fig.~\ref{Fig5}~(b), the variation of measurement precisions $\Delta \kappa$ versus $\kappa$ for $N=20$ is shown.
According to our result, we find that the optimal measurement precision $\Delta \kappa_{{{min}}}$ occurs near $\kappa T =0$.

Further, to confirm the dependence of measurement precision $\Delta \kappa$ on particle number $N$, we numerically calculate the optimal measurement precision versus particle number $N$ in the range of $\kappa T\in [-0.1\pi,0.1\pi]$, as shown in Fig.~\ref{Fig5}~(a).
According to the fitting result, we find the optimal measurement precision of LSV parameter $\kappa$ is ${{\textrm{In}}}(\Delta \kappa_{{min}}T)=-0.8\textrm{In}(N)+0.43$.
Our study indicates that the beam recombination via reversed sweeping through QPTs also can make the measurement precision of the LSV parameter $\kappa$ beyond SQL.
\section{Robustness against imperfections \label{Sec5}}
%
In this section, we study the robustness of the multimode many-body quantum interferometry.
In practical experiments, many imperfections can limit the final measurement precision.
Here, we discuss two imperfections that mainly exist in the scheme of driving through QPTs: (i) the non-adiabatic effect in the initialization and recombination steps, and (ii) the detection noise in the measurement step.
\subsection{Influence of nonadiabatic effect \label{Sec5.1}}
\begin{figure}[!htp]
  \includegraphics[width=1\columnwidth]{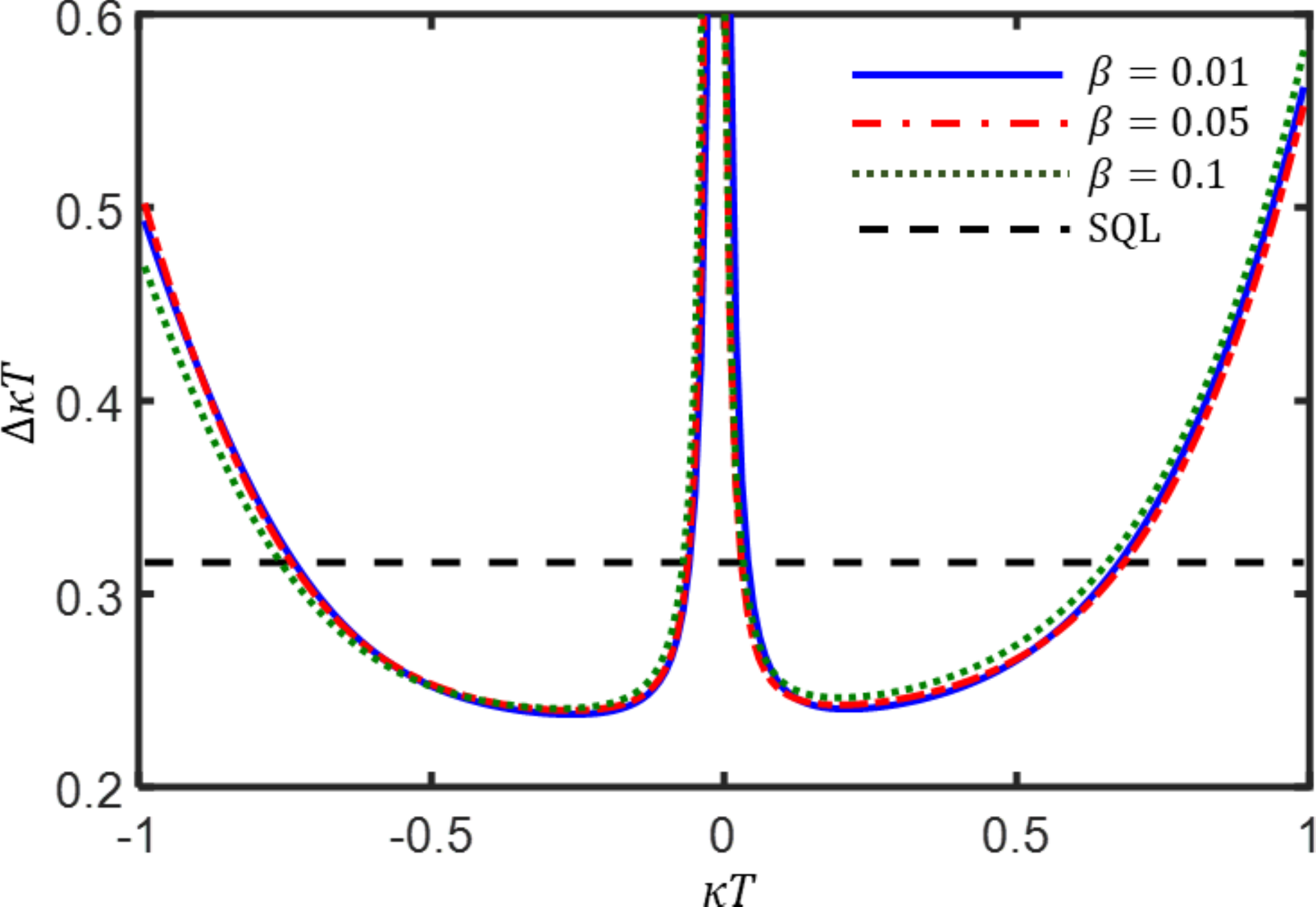}
  \caption{\label{Fig6}(color online).
  The measurement precision $\Delta \kappa$ versus $\kappa$ for different sweeping rate $\beta$ under $N=10$.
  Here, $c_{2}=-1$, $q_0 =-q_f=3$, $T=1$ and  $N=10$.}
\end{figure}
Firstly, we consider the non-adiabatic effect.
To realize the proposal of driving through QPTs perfectly, the sweeping process should be adiabatic.
However, non-adiabatic effect always exists in practical experiments.
In general, the adiabaticity of the driving process can be characterized by the sweeping rate $\beta$~\cite{Ref61}.
If the sweeping rate $\beta$ is sufficiently small, the adiabatic evolution can be achieved.
To confirm the influence of non-adiabatic effect on the measurement precision, the variation of $\Delta\kappa$ with $\kappa$ for different sweeping rate $\beta$ is shown in Fig.~\ref{Fig6}.
It can be observed that the measurement precision of $\Delta\kappa$ becomes worse as $\beta$ increases, but the measurement precision $\Delta\kappa$ can still beat the SQL when the sweeping rate $\beta$ is moderate.
\subsection{Robustness against detection noise \label{Sec5.2}}
\begin{figure}[!htp]
  \includegraphics[width=1\columnwidth]{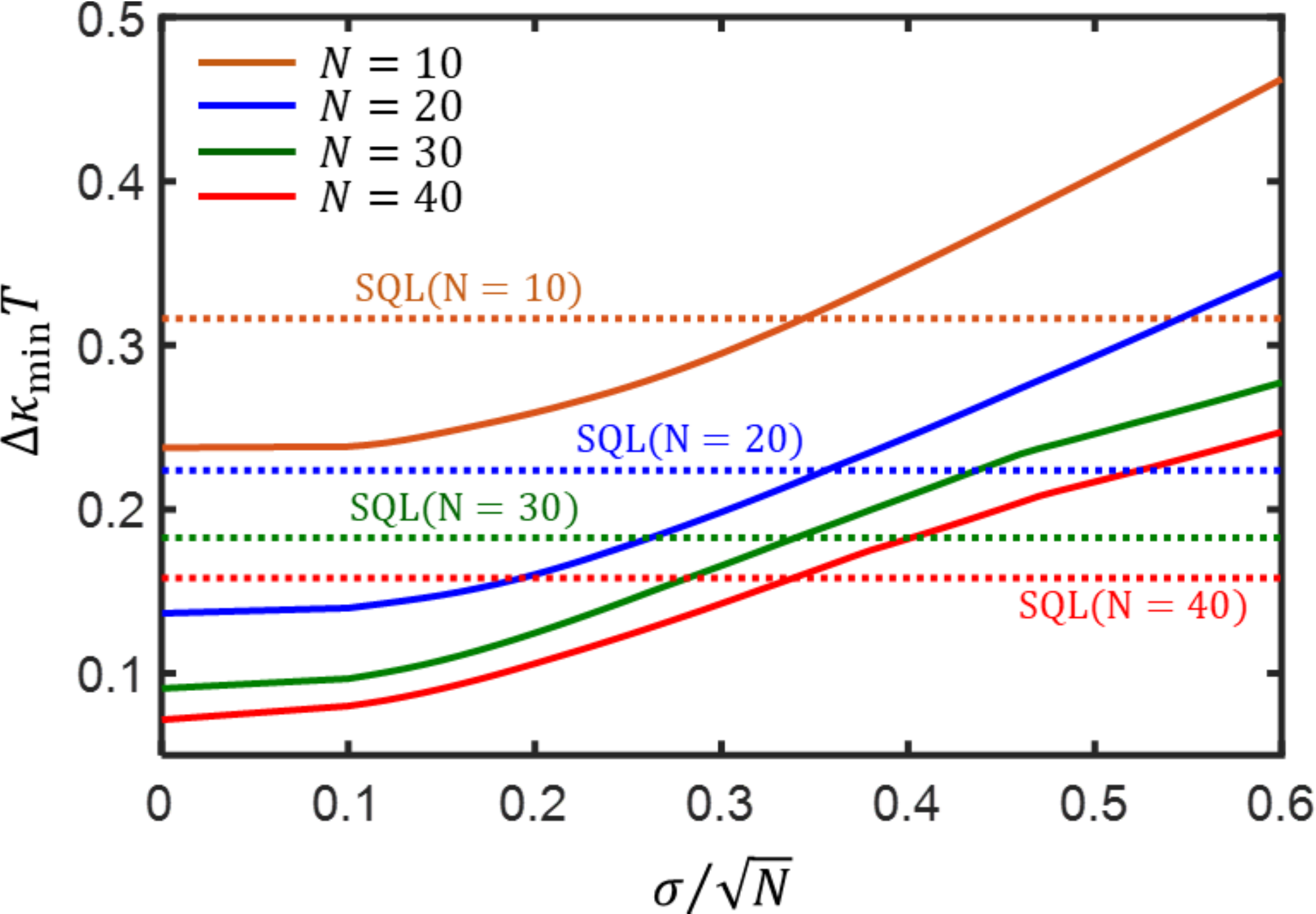}
  \caption{\label{Fig7}(color online).
  The optimal measurement precision $\Delta{\kappa}_{\textrm{min}}T$ versus detection noise $\sigma$ for
  $N=10, 20, 30, 40$.
  The dotted lines denote the corresponding SQL, i.e., $1/\sqrt{N}$.
  Here, $c_{2}=-1$, $q_0 =-q_f=3$, $T=1$ and  $\beta=0.05$.}
\end{figure}
Here, we study the influence of detection noise by considering the additional classical noise in the measurement process.
In an ideal situation, the population measurement on the final state can be rewritten as $\langle\hat{N}_{0}\rangle=\sum_{N_0}{P(N_0|\kappa)}N_0$, where ${P(N_0|\kappa)}$ is the ideal conditional probability which obtain measurement result $N_0$ with a given $\kappa$.
However, in realistic experiments, the detection noise can limit the measurement precision of frequency.
For an imperfect detector with Gaussian detection noise~\cite{Ref61,LucaPezze2013,DMStamperKurn2013,SPNolan2017}, the population measurement becomes
\begin{equation}\label{Eq:detection_noise}
\langle\hat{N}_{0}\rangle=\sum_{N_0}\tilde{P}(N_0|{\sigma})N_0,
\end{equation}
with
\begin{equation}\label{Eq:detection_noise_p}
{\tilde{P}(N_0|{\sigma})}=\sum_{\tilde{N}_0} A_{\tilde{N}_0}{e^{-(N_0-\tilde{N}_0)^2/2{\sigma}^2}}{P({N}_0|{\kappa})}.
\end{equation}
the conditional probability depends on the detection noise $\sigma$.
Here, $A_{\tilde{N}_0}$ is a normalization factor.
%
To study the influence of the detection noise, the optimal measurement precision $\Delta{\omega}_{\textrm{min}}$ versus the detection noise $\sigma$ for different particle number $N$ is shown in Fig.~\ref{Fig7}.
For $N=10,20,30,40$, our numerical results indicate that the measurement precision can still beat the SQL when $\sigma\leq 0.35 \sqrt{N}$.
There is no need to use single-particle resolved detector for measurement when $N$ is modestly large.
Thus our proposal is robust against to the detection noise.
\section{Summary and discussion \label{Sec6}}

In summary, we have presented an entanglement-enhanced multimode quantum interferometry for testing the LSV effect.
The LSV effect for a bound electron system results in an energy shift, which is the directional dependent interaction (from the SME).
Based upon the proposed scheme, we study the ultimate measurement precision bound for the LSV parameter $\kappa$ via individual atoms and entangled atoms.
We find that the ultimate measurement precision bound depends on both the spin length $F$ and the total atom number $N$.
Larger spin length $F$ and total atomic number $N$ may offer better measurement precision.
Especially, if the input state is the $N$-atom multimode GHZ state, the measurement precision of $\kappa$ can attain the Heisenberg limit, i.e., $\Delta \kappa \propto 1/(F^2N)$.

Moreover, we study the LSV test via the experimentally accessible three-mode interferometry with Bose condensed spin-$1$ atoms~\cite{Ref40}.
The three-mode interferometer is available in experiments with spin-1 atoms~\cite{YZou2018,Luo2017,SGuo2021}.
By selecting suitable input states and recombination operations, the LSV parameter $\kappa$ can be extracted via population measurement.
Especially, the measurement precision of the LSV parameter $\kappa$ can beat the SQL and even approach the Heisenberg limit via SMD~\cite{Linnemann2016} or driving through QPTs~\cite{ZZhang2013,YZou2018,Luo2017,SGuo2021,Luo2017}.
To realize the entanglement-enhanced test for local LSV with the three-mode interferometer in experiments, the precise control of system parameters and the implementation of adiabatic sweeping process are both necessary.
Owing to the well-developed techniques in quantum control, the entangled states can be generated in $^{87}$Rb condensate via SMD or adiabatically driving through QPTs with state-of-the-art techniques.

According to the relationship between the parameters $\kappa$ and $C^{(2)}_0$~\cite{Ref31}, the measurement precision of parameter $C^{(2)}_0$ can be improved the same orders of magnitude.
According to Eq.~(\ref{Eq2}), the relationship between the parameters $\kappa$ and $C^{(2)}_0$ reads as
\begin{eqnarray}\label{Eq30}
 \frac{\kappa}{2\pi}=\frac{\Delta E/(hC^{(2)}_0)}{\Delta (j_{z}^2)}C^{(2)}_0,
\end{eqnarray}
here $h$ is the Planck constant,
$\Delta E$ and $\Delta (j_{z}^2)$ denote the energy deviation and angular momentum fluctuation for the experimentally selected states, respectively.
One can obtain $\Delta E$ via calculating the matrix element of operator $T_0^{(2)}$, which is proportional to $m_F^2$ and the reduced matrix element $\langle F|T^{(2)}|F\rangle$ simultaneously.
For the $^{87}\textrm{Rb}$ atoms, a rough estimate using the method of Refs.~\cite{Ref26,Ref31} gives $\Delta E/(hC^{(2)}_0)=8.6 \times 10^{15}$ Hz and $\Delta (j_{z}^2)=1$ for the $m = 1$ and $m=0$ states.
Compared with the measurement precision reported in Ref.~\cite{Ref31} which just asymptotically approaches the SQL via observable measurement, our scheme can be expected near the Heisenberg limit via population measurement.
Thus, also assuming a year-long measurement with $10^4$ atoms, the measurement precision of parameter $C^{(2)}_0$ is about $2$ orders magnitude higher than the result reported in Ref.~\cite{Ref31}.
Compared with the LSV tests via two fully entangled Yb$^{+}$ ions~\cite{Ref25} with the parameter $C^{(2)}_0$ will be bounded at the level of $10^{-23}$ for year-long measurement, our scheme can also be used for improving the corresponding measurement precision by utilizing multi-particle entangled states.

Our proposal may open up a feasible way for a drastic improvement of the LSV tests with atomic systems and also provide an experimentally accessible scheme to realize entanglement-enhanced test for LSV.
Besides, it may be used for testing some fundamental theories, such as field theory, string theories, Einstein-aether theories, and quantum gravity.

\acknowledgements{This work is supported by the National Key Research and Development Program of China (Grant No. 2022YFA1404104), the National Natural Science Foundation of China (Grant No. 12025509, 11874434), and the Key-Area Research and Development Program of GuangDong Province (Grant No. 2019B030330001). J. H. is partially supported by the Guangzhou Science and Technology Projects (202002030459)}.
%


%
\onecolumn
\appendix
\setcounter{equation}{0}
\renewcommand{\theequation}{A\arabic{equation}}
\section*{Details of the numerical simulation}
Here, we introduce the details of our numerical simulation, we start from all $N$ atoms in state $|{1,{m_{F}=0}}\rangle$, i.e., $|{\Psi_0}\rangle=|{0,N,0}\rangle$.
Due to the conservation of magnetization $N_1=N_{-1}$ under spin-exchange collisions, the dynamical evolution will be constrained in the zero magnetization subspace of
$\{|k,N-2k,k\rangle\}$, with $k,N-2k,k$ representing number of atoms in the $m_F = +1, 0, -1$ components, respectively.
For the scheme of LSV test via spin-mixing dynamics, the matrix element of Hamiltonian $\hat{H}_{SMD}$ in this basis is given by
\begin{eqnarray}\label{Eq:matrix1}
[\hat{H}_{SMD}]_{k',k}&\equiv&\langle k',N-2k',k'|\hat{H}_{SMD}|k,N-2k,k\rangle \nonumber \\
&=&\chi k \sqrt{(N-2k+1)(N-2k+2)} \delta_{k',k-1}\nonumber \\
&+&\chi(k+1)\sqrt{(N-2k)(N-2k-1)} \delta_{k',k+1}\nonumber \\
\end{eqnarray}
For the scheme of LSV test via driving through quantum phase transitions, the matrix element of Hamiltonian $\hat{H}_{QPT}$ in this basis is given by
\begin{eqnarray}\label{Eq:matrix2}
[\hat{H}_{QPT}]_{k',k}&\equiv&\langle k',N-2k',k'|\hat{H}_{QPT}|k,N-2k,k\rangle \nonumber \\
&=&[\frac{c_{2}}{N}k(2N-4k-1)-q(N-2k)]\delta_{k',k}\nonumber \\
&+&\frac{c_{2}}{N} k \sqrt{(N-2k+1)(N-2k+2)} \delta_{k',k-1}\nonumber \\
&+&\frac{c_{2}}{N}(k+1)\sqrt{(N-2k)(N-2k-1)} \delta_{k',k+1}\nonumber \\
\end{eqnarray}
To simulate dynamical process, we expand the wave function at time $t$ in the basis $\{|k,N-2k,k\rangle\}$ as
\begin{eqnarray}\label{Eq:dynamical_evolution}
|{\Psi(t)}\rangle=\sum_{k=0}^{N/2}C_{k}(t)|k,N-2k,k\rangle.
\end{eqnarray}
According to the Schrodinger equation $i\partial_{t} |{\Psi(t)}\rangle=H|{\Psi(t)}\rangle$, one can obtain the coupled equations for the superposition coefficients $C_{k}(t)$.
For the Hamiltonian $\hat{H}_{SMD}$, the coupled equations are
\begin{eqnarray}\label{Eq:ck1}
i\partial_{t}C_{k}^{SMD}(t)&=&\chi k \sqrt{(N-2k+1)(N-2k+2)} C_{k-1}^{SMD}(t)\nonumber \\
&+&\chi(k+1)\sqrt{(N-2k)(N-2k-1)} C_{k+1}^{SMD}(t)\nonumber \\
\end{eqnarray}
For the Hamiltonian $\hat{H}_{QPT}$, the coupled equations are
\begin{eqnarray}\label{Eq:ck2}
i\partial_{t}C_{k}^{QPT}(t)&=&[\frac{c_{2}}{N}k(2N-4k-1)-q(N-2k)]C_{k}^{QPT}(t)\nonumber \\
&=&\frac{c_{2}}{N} k \sqrt{(N-2k+1)(N-2k+2)} C_{k-1}^{QPT}(t)\nonumber \\
&+&\frac{c_{2}}{N}(k+1)\sqrt{(N-2k)(N-2k-1)} C_{k+1}^{QPT}(t)\nonumber \\
\end{eqnarray}
This series of coupled equations can be solved numerically  and one can obtain any expectations for the interested observables.
\end{document}